\documentclass{iaus}
\usepackage{graphicx}

\title[AGB stars]
{Mass loss on the Asymptotic Giant Branch}

\author[A.A. Zijlstra]   
{Albert A. Zijlstra}

\affiliation{Jodrell Bank Centre for Astrophysics, School of Physics \&\
  Astronomy, University of Manchester, P.O. Box 88, Manchester M60 1QD
\break
email: a.zijlstra@manchester.ac.uk}

\pubyear{2006}
\volume{234} 
\pagerange{1--8}
\date{}
\jname{Proceedings Planetary Nebulae in Our Galaxy and Beyond}
\editors{M.J. Barlow \&\ R.H. Mendez, eds.}

\begin{document}

\maketitle

\begin{abstract}
Mass loss on the Asymptotic Giant Branch provides the origin of planetary
nebulae.  This paper reviews several relevant aspects of AGB
evolution: pulsation properties, mass loss formalisms and time
variable mass loss, evidence for asymmetries on the AGB, binarity, ISM
interaction, and mass loss at low metallicity. There is growing
evidence that mass loss on the AGB is already asymmetric, but with
spherically symmetric velocity fields. The origin of the rings may be
in pulsational instabilities causing mass-loss variations on time scales of
centuries. 
\end{abstract}

\firstsection 
\section{Introduction}

The Asymptotic Giant Branch (AGB) contains two separate phases. During
the early AGB, stars experience quiescent shell burning around the
inert but growing carbon--oxygen core. Later, quiescent hydrogen
burning is punctuated by run-away helium flashes \cite[(Herwig
2005)]{Herwig2005} followed by short phases of (lower-luminosity)
quiescent helium burning. The helium flashes are called thermal pulses
(TPs). During the TP-AGB phase, the star begins to experience strong
pulsations, and increasing mass loss. Evolution proceeds from
semiregular (SR) to Mira and eventually to OH/IR star, with mass-loss
rates simultaneously increasing from typically $10^{-6}\,\rm
M_\odot\,yr^{-1}$ for Miras, to $10^{-4}\,\rm M_\odot\,yr^{-1}$ for
OH/IR stars. The latter show strong self-obscuration by the
circumstellar envelope. Carbon-rich stars show a parallel sequence but
with lower V-band amplitudes, and higher circumstellar extinction for
the same mass-loss rates.

The mass loss reaches values much higher than the nuclear burning
rate. The hydrogen envelope therefore diminishes much faster than that
the core grows. Once the envelope mass falls below a critical value
($\sim 0.02\,\rm M_\odot$), the photosphere begins to collapse,
mass-loss rates reduce and the star enters the post-AGB phase. In this
way, the mass loss causes a very sudden end to stellar evolution
\cite[(e.g. Willson 2000)]{Willson2000}.

In this paper, aspects of AGB evolution related to the mass loss will
be reviewed.

\section{Pulsations}
Pulsation properties evolve strongly during the TP-AGB.  The
semiregulars have periods of typically 50--150\,days, small amplitudes
and (as the name implies) have irregular light curves. Mira variables
have periods of 150--500\,days, amplitudes in excess of 2.5\,mag in V,
and stable light curves. OH/IR stars show periods of 500--2000 days.
The amplitude is largest at optical wavelengths. The stellar energy
distribution peaks in the infrared, and the bolometric amplitude is
much less than in the V-band, rarely more than one magnitude.  The
large optical amplitude is caused by variable molecular opacity,
mainly due to TiO and VO bands. The classification can show
imperfections. OH/IR stars tend not to be observable in the V-band,
and are often classified as SR because the amplitude is only measured
at K. Stars with higher C/O ratio show reduced abundances of
oxygen-rich molecules, and can therefore also show lower amplitudes.

The pulsation equation is given by:

\begin{equation}
\begin{array}{lrllll}
 \log P &=& 1.949 \log R-0.9 \log M - 2.07 & 
             \hbox{fundamental mode \cite[Wood 1990)]{Wood90}} \\
 \log P &=& 1.5\log R\ \ \ -0.5 \log M + \log Q &  
              \hbox{first overtone \cite[(Fox \&\ Wood 1982)]{FW82}},
\end{array}
\end{equation}
\noindent
where in the latter case the pulsation constant $Q \approx 0.04$; the
period $P$ is in days and the radius $R$ and mass $M$ are in solar
units. Mira variables are fundamental-mode pulsators, whilst SR
variables show overtone pulsations. Fundamental-mode models including
turbulent viscosity \cite[(Olivier \&\ Wood 2005)]{Olivier2005} reproduce
Mira light curves fairly well.

A well-defined Mira period--luminosity relation was found from stars
in the LMC. The Macho, Ogle and Sirius surveys have shown that this 
relation is only one of several parallel
sequences. Magellanic Cloud stars show five sequences \cite[(e.g. Kiss
\&\ Bedding 2003, Ita et al. 2004)]{Kiss2003, Ita2004}, two of which
are identified as overtones, one is the Mira fundamental mode, one is
an RGB sequence tracing semi-detached binaries, and the last sequence shows
periods much longer than the fundamental mode which are still
unexplained \cite[(Wood et al. 2004)]{Wood2004}.

Long-term monitoring of Mira variables, mostly by amateur astronomers,
has given light curves covering, in some cases, several
centuries. These extended light curves have shown that most Miras are
stable, but a number of stars show evolving periods. The best known
case is that of R Hya, for which historical records \cite[(Zijlstra et
al. 2020]{Zijlstra2002} indicate a continuous decline from about 495
days before 1800, to 385 days in 1950. The period has remained stable
since 1950, and the earlier longer period also may have been
stable. \cite{Templeton2005} has analyzed the full set of AAVSO
records available, and out of 547 Mira variables find 8 cases of long-term
period changes, at 6$sigma$ significance. The most dramatic case is T
UMi, which showed a constant period for over 60 years, before a sudden,
sharp decrease in period by 25 per cent over the next 20 years.

\cite{Zijlstra2002a} define three types of period instability: sudden
changes as in T UMi, continuous changes as in R Hya, and meandering
periods. The latter appears to be common among the longest period Mira
variable, and involve variations up to 10 per cent on time scales of
$\sim 50\,$years.

Several explanations exist for the period changes. Thermal pulses
involve large, sudden variations in luminosity (and therefore radius
and period). This fits stars such as T UMi well, where there was no
pre-indicator of a developing instability.  The rate of thermal pulses
is of the order of 0.5 per cent per century, or about $\sim 2$ per
century among the known Miras. The second possiblity is a non-linear
pulsation \cite[(Olivier \&\ Wood 2005)]{Olivier2005}, where the
change in amplitude causes a change in period \cite[(Ya'ari \&\
Tuchman 1996)]{Yaari1996}. Following a large amplitude change,
e.g. the onset of pulsation, they find a large period decrease of the
fundamental mode over a time scale of centuries. The observed
correlation between period and amplitude changes \cite[(e.g Zijlstra
\&\ Bedding 2002)]{Zijlstra2002a} provides some support for this
model. \cite{Icke1992} have pointed out that the surfaces of
long-period stars are susceptible to weak chaos. \cite{Zijlstra2004a}
discuss strongly variable periods among SC stars, where the C/O ratio
is very close to unity and the chemical equilibrium is highly sensitive
to temperature changes: they suggest that in SC stars, rapid changes
in molecular abundances may trigger weakly chaotic period changes.

\section{Mass loss}
\subsection{Proposed relations}

AGB mass loss is a multi-step process. The pulsations extend the
atmosphere, and allow molecules to form. At large distances the
temperature drops below the condensation temperatures and a number of
solids form: MgSiO-bearing grains in O-rich environments, and SiC and
amorphous carbonaceous particles around carbon-rich stars. Radiation
pressure pushes the grains out, and by collisions carries the gas with
it. Pulsation can support mass-loss rates $\sim 10^{-7}\,\rm
M_\odot\,yr^{-1}$, but the radiation pressure on dust (and molecules)
greatly increases the mass loss. The difference corresponds to the
case A and case B mass-loss regimes of \cite{Winters2000}.

Dynamical models of the extended envelopes  do not yet convincingly
predict mass-loss rates. Instead, formulistic mass-loss rates are
used. The following relations are in general use:

\begin{equation}
\begin{array}{rccll}
 (i) & \log \dot M  & = & a\, P + b & \qquad \rm 
             \cite[Vassiliadis\ \&\ Wood 1993]{Vassiliadis1993} \\
 (ii) & \dot M      & = & a\, M_{\rm i}^{-2.1}\,  L^{3.1}\, R\, M^{-1}  &  
                     \qquad \rm \cite[Bloecker\ 1995]{Bloecker1995} \\
 (ii) & \dot M      & = & a\, L^{2.47}\, T^{-6.8}\, M^{-1.95} &
    \qquad \rm \cite[Wachter\ et\ al.\ 2002]{Wachter2002} \\
 (iv) & \dot M & = & a\, L^{1.05}\, T^{-6.3} & \qquad \rm
             \cite[van\ Loon\ et\ al.\ 2005]{VanLoon2005}.
\end{array}
\end{equation}

\noindent
Relation (ii) is derived by fitting the Bowen pulsation models to observed
initial--final mass relations; (iii) is derived for carbon-rich stars
only, using dynamical models, whilst (iv) comes from mass-loss
determinations in the LMC, including both AGB and RSG stars, and is
observationally the best established.

A very strong dependence on stellar temperature is indicated. However,
the optical spectral type used to assign a temperature measures the
$\tau=1$ layer in the molecular envelope, far above the real stellar
surface: it underestimates both the surface and the effective
temperature. 
The strong dependence is also deceptive. Using $L
\propto R^2 T^4$, relations (iii) and (iv) give $ \dot M \propto
T^{3.6}$ and $T^{-2.3}$, respectively. The first one is physically
unlikely, and \cite{Wachter2002} may overestimate the effect of
luminosity. The Bloecker relation suffers from the same
problem. Relation (iv) excludes mass as a relevant parameter, because
this parameter is not known for the LMC stars. However, we may adopt
the suggestion that the mass loss depends on the surface binding
energy, $M/R$.  In this case, relations (iii) and (iv) suggest that
the total dependence is along the lines of

\begin{equation}
 \dot M \sim \left( \frac{M}{R} T \right)^\alpha L^\gamma, 
\end{equation}

\noindent with $\alpha \sim -2$; the luminosity dependence is
likely  weaker, with $\gamma =\sim 1$. 

In addition to the surface parameters, the mass loss is also affected
by the period, since the atmosphere extension is pulsation-driven.
Relation (i) is the only one to include the pulsation explicitly. But
the other relations also implicitly do so, via the pulsation equation.
However, they predict a period dependence which is much weaker than
found for (i).  An  example to constrain the period dependence is
given by the evolution of R Hya, because its decreasing period over
the past 200 years was accompanied by a strong decrease in mass loss
\cite[(Zijlstra et al. 2002)]{Zijlstra2002}. Its evolution fits
relation (i) well, but is not be reproduced by the other relations.

Overall, the predictive value of the proposed relations is still
limited. Important parameters, such as composition and metallicity,
are not or not fully included. The case A and B regimes may show
different dependences on stellar parameters.

\subsection{Time-variable mass loss}

Mass loss is known to vary on several time scales. Over $10^6\,\rm
yr$, the slow increase in the absolute magnitude of $\sim
1\,\rm mag$ will be accompanied by an increase of $\dot M$
\cite[e.g. Vassiliadis \&\ Wood 1993]{Vassiliadis1993}. This slow
evolution is interrupted by the thermal pulses, which cause a
sharply enhanced mass loss event, occuring every $10^4$ to
$10^5$ yr, depending on the core mass. Mass loss variations on a time
scale of $10^2$--$10^3$\,yr are implied by the multiple rings seen
around many PNe. Finally, extinction variations in both carbon-rich
and oxygen-rich stars indicate changes within decades.

 Mass loss spikes during the TP are the accepted explanation for the
detached shells seen around TT Cyg, U Cam and DR Ser, among others
\cite[(e.g. Olofsson et al. 2000)]{Olofsson2000}; the thin shells can be
explained by a brief burst of 100-fold stronger mass loss, sweeping up
the previous, weaker and slower wind \cite[(Sch\"oier et
al. 2005)]{Schoier2005}. The TP is followed by a subsequent phase of
quiescent helium burning when $L$ is a factor of 3 lower than during
the hydrogen burning phase, with a  weaker wind.

The multiple, thin rings seen around PNe \cite[(Corradi et
al. 2004)]{Coradi2004} still lack a commonly accepted
explanation. First, an instability involving the dust driving was
suggested by \cite{Simis2001}: dust formation leads to increased dust
density, until dust-gas coupling empties the dust shell. This leads to
mass-loss episodes at the correct time scales. However, this
instability is not correlated between different directions, and does
not lead to spherical structures. Cooling due to existing dust
strongly enhances the local dust formation rate, leading to
near-chaotic mass loss \cite{Woitke2005}.  Second, a nuclear-burning
instability was suggested by \cite{VanHorn2003}. Third, the period
instability found for long-period Miras, interpreted as due to a
non-linear pulsation, could cause the rings \cite[(Zijlstra \&\
Bedding 2002]{Zijlstra2002a}, via period-induced mass-loss variations.
The hypotheses need to explain the slight irregularity in the ring
spacings, and the decrease in spacings for rings closer to the star
(i.e. ejected later).  The model involving the non-linearity in the
pulsations can explain both. However, the issue is still far from
settled.

The decadal (extinction) variations are most easily explained by the
dust clumping as found in the Woitke models.
 Extinction variations are mainly seen in
carbon-rich stars \cite[(e.g. Whitelock et al. 2006)]{Whitelock2006}
but a few oxygen-rich stars also show variations due to circumstellar
extinction, most noticeably L$_2$ Pup \cite[(Bedding et
al. 2002)]{Bedding2002}.

\section{Morphology}

Strong asymmetries are present in the morphologies of almost all
post-AGB envelopes. These structures must have their origin during the
AGB mass loss.  But remarkably little evidence for the precursors of
the post-AGB morphologies has been found. In fact, the current
consensus is that AGB envelopes are spherically symmetric, and that
asymmetries form during a run-away process at the very tip of the AGB.
The argument that such a fundamental change
happens during a phase which is unobservable only by its brevity,
is far from convincing.

Two observations provide  evidence for spherical AGB shells. One
comes from the OH maser profiles of OH/IR stars,
with strong peaks at the extreme velocities.  For asymmetric
envelopes, the emission at intermediate velocities becomes stronger
and multiple peaks may develop \cite[(Chapman 1988)]{Chapman1988}.  The
second observation is that the outer structures around PNe and
post-AGB stars are often round, especially so for the rings. The
regular detached shells around, e.g., TT Cyg should also be mentioned.

However, both can be questioned. Maser emission cannot be used to
determine a density structure. Instead it is an exponential function
of the column density at constant velocity, in competition between all
directions.  OH profiles are far more dependent on the velocity
structure than on the densities.  An asymmetric envelope with a
spherically symmetric velocity structure will give rise to symmetric profiles.

The rings appear to provide stronger evidence. However, any mass-loss
changes leading to such rings are likely to be accompanied by changes
in the expansion velocity. The rings then become the swept-up
interface between the more massive, faster wind and the slower, weaker
wind: \cite{Zijlstra2001} show that the swept-up shell will move at
constant velocity, with the location at anyone time given by:

 \begin{equation}
 r_{\rm ring} = v_{\rm ring} t = V_{\rm slow} {1 \over {1 - \mu / \xi}}
  \left[ 1 - \mu + (\xi - 1) \sqrt{\mu / \xi} \right],
\end{equation}
 
\noindent where 
 
  $$\mu = {{{\dot m}_{\rm fast}} \over {{\dot m}_{slow}}} \quad ; \qquad
  \xi = {{V_{\rm fast}} \over {V_{\rm slow}}}. $$

If the two winds are self similar, and specifically if they show
spherically symmetric velocities and show the same ratio of mass-loss
rates in all directions, $v_{\rm ring}$ will be direction independent.
The swept-up interface will be spherical, even if the underlying mass loss
is not.

Thus, in both cases there is strong evidence for spherically
symmetric{\it velocity fields}, but the mass loss itself can be
asymmetric.

\begin{figure}
\begin{center}
 \includegraphics[width=5cm, clip=]{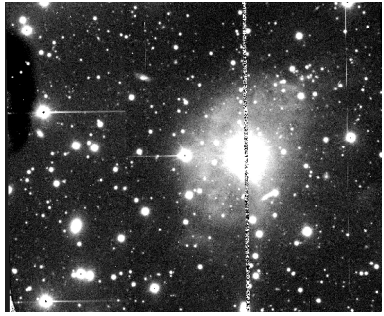}
 \caption{NTT V-band image of RZ Sgr, showing its bipolar reflection nebula}
\label{rzsgr}
\end{center}
\end{figure}

There is in fact growing evidence for bipolar shells on the AGB. The
bipolar envelope of X Her was mapped by \cite{Nakashima2005}.  The
reflection nebula around RZ Sgr \cite{Whitelock1994} is bipolar in our
NTT image (Fig. \ref{rzsgr}).  Linear polarization is present in
integrated photometric data of some OH/IR stars \cite[(Jones \&\ Gehrz
1990)]{Jones1990}.  Finally, the core of IRC+10\,216 shows several
components, possibly in a bipolar configuration \cite[(Osterbart et
al. 2000)]{Osterbart2000}. (The last two points can also be expained by
Woitke-type blobs \cite[(Woitke 2006)]{Woitke2006}, however.)

\section{Binarity}

Whereas binarity is now thought to be very common among post-AGB stars
and PNe, and the likely cause of the asymmetries of the envelopes, very
little is known about binarity among AGB stars. The extreme brightness
and low temperature of the AGB star make a binary comnpanion difficult
to detect, except where the companion interacts with the extended
atmosphere or the wind. Direct evidence for binarity among
AGB stars will be needed to make evolutionary connections between AGB and
post-AGB stars.

A distinction can be made between four different types of interaction. 
\begin{enumerate}
\item Common envelope systems, where the companion triggers the mass
loss. These system will  rarely reach the AGB, as the common
envelope is likely to develop already on the RGB. Orbital separations
are of the order of 1\,AU. Perhaps 10 per cent of 'proto planetary
nebulae' may derive from such systems, although fewer would evolve
into PNe.

\item Wider systems, where the geometry of the mass loss is affected,
but not the mass-loss rate itself. These systems may include symbiotic
stars and the RV Tau progenitors. Orbital separations are
a few to  tens of AU, and the systems exhibit
circumbinary disks and accretion. These may account for $\sim 25$ per
cent of the PN birth rate. The companion may 
ionize part of the stellar wind, as in OH 231.8+4.2.

\item Systems with only minor effects on the mass loss
geometry. Typical separations are $\sim 100\,$AU. The
geometry may show  the movement of the mass-losing star, as 
in the spiral nebula around AFGL 3068 \cite[(Mauron \& Huggins 2006,
see also Morris et al, these proceedings)]{Mauron2006}. Mira itself 
has a white dwarf companion $\sim 100$\,AU away.

\end{enumerate}

\section{ISM interaction}

The study of the interaction between the AGB wind and its local ISM
was pioneered by \cite{Villaver2002}. Previously, the interaction had
only been considered for old, ionized PNe. The interaction begins in
fact with the onset of the AGB wind, and leads to a swept-up shell
surrounding the AGB star. Once the swept-up mass exceeds the mass lost
in the wind, the shell slows down and eventually becomes
near-stationary. The typical radius of this shell is 1--2 pc,
depending on local conditions. An example of such a shell, called a
'wall', was found by \cite{Zijlstra2002b}. The wall may wipe out any
memory of long-term mass-loss fluctuations. Once the PN fast wind
reaches the outer wall, the PN rebrightens; this increases the PN life
time.

If there is a significant proper motion of the star with respect to
 the ISM, the wall becomes one-sided and at higher velocities, becomes
 a bow shock. This is modelled by \cite{Wareing2006}, based on the
 IPHAS morphology of the PN Sh2-188. The Spitzer image of R Hya
 (Speck, these proceedings) show another example of such a bow shock.)
 Evolutionary calculations for a large range of parameters are
 presented by Wareing (these proceedings).

\section{Metallicity}

The effect of metallicity on AGB mass loss is only just beginning to
be explored. \cite{Bowen1991} first showed that for [Fe/H]$<-1$,
dust-driven winds fail and the mass loss becomes pulsation-driven
instead. This significantly reduces the mass-loss rates. Using their
formalisms, \cite{Zijlstra2004} find that the final (white dwarf) mass
becomes higher at very low metallicity, and may reach the Chandresekhar
limit. This does not happen in the models of \cite{Herwig2004} where
core-dredge-up limits the core mass to $\approx 1\,\rm M_\odot$.

\begin{figure}
\begin{center}
 \includegraphics[width=10cm]{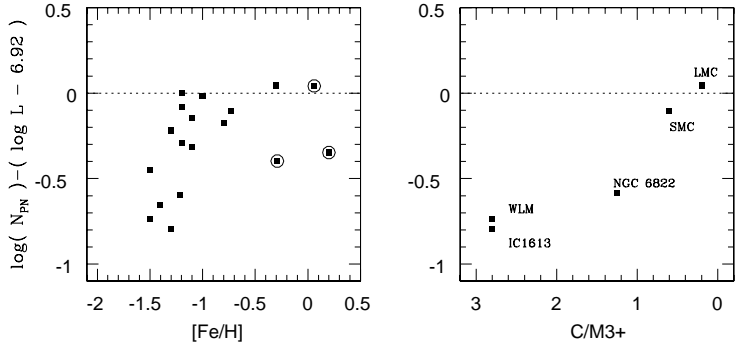}
 \caption{ Left: The ratio between number of planetary nebulae
and luminosity of the parent stellar population ($y$-axis), as
function of metallicity ($x$), for nearby galaxies, from
\cite{Magrini2003}. Encircled points indicate spiral galaxies affected
by internal extinction. Right: The same ratio, as function of the ratio of
carbon over M-type stars on the AGB, a tracer of the metallicity of
the AGB population}
\label{feh}
\end{center}
\end{figure}

Mass-loss rates in the LMC are as high as those in the Galaxy
\cite{VanLoon1999}. However, it is possible that these values are
reached at higher luminosities than for Galactic stars. (The
metallicity of the LMC is comparable to the outer Galactic disk.)
Evidence for reduced mass-loss rates at much lower metallicity comes
mainly from the specific frequency of PNe, which is significantly
lower in metal-poor dwarf galaxies compared to more metal-rich
systems. This is illustrated in fig.  \ref{feh}.

Expansion velocities will be much lower at lower metallicity, limited
by the dust-to-gas ratio. Evidence is sparse, as expansion velocities
outside of our Galaxy have only been measured for some LMC OH/IR
stars. These indeed indicate a significantly lower expansion velocity in
the LMC \cite[(Marshall et al. 2004)]{Marshall2004}.

Dredge-up of carbon will lead to higher C/O ratios at lower
metallicity, because of the lower oxygen abundance. This leads to the
counter-intuitive effect that carbon molecules are more abundant at
lower metallicity. The C$_2$H$_2$ infrared bands were indeed found to
be very strong in carbon stars in the Magellanic Clouds
\cite[(e.g. Matsuura et al. 2005)]{Matsuura2005}. This has
consequences for dust formation: amorphous dust may form easily around
low metallicity carbon stars, while minerals such SiC and MgS do not,
ginvig a mineral-poor or 'soft' dust.
\cite{Sloan2006} and \cite{Zijlstra2006}, using Spitzer spectra,
indeed find that the SiC and MgS features are weaker in the Magellanic
Clouds. Oxygen-rich stars can only form dust from metal-dependent
minerals, and their dust formation efficiency is expected to be
strongly suppressed at low metallicity. Mass loss and dust formation
at low metallicity is therefore expected to be dominated by carbon stars.



\begin{thebibliography}{}

\bibitem[Bedding et al. (2002)]{Bedding2002} {Bedding T.R., Zijlstra
    A.A., Jones A., Marang F., Matsuura M., Retter A., Whitelock P.A.,
    Yamamura I.}, 2002, MNRAS, 337, 79

\bibitem[Bloecker (1995)]{Bloecker1995}
     {Bloecker T.,} 1995,
     A\&A, 297, 727

\bibitem[Bowen \&\ Willson (1991)]{Bowen1991}
     {Bowen G.H., Willson L.A.} 1991,
     ApJ, 375, L53

\bibitem[Chapman (1988)]{Chapman1988}
      {Chapman J.M.}, 1988, MNRAS, 230, 415

\bibitem[Corradi et al. (2004)]{Corradi2004}
      {Corradi R.L.M., S\'anchez-Bl\'azquez P., Mellema G., 
       Giammanco C., Schwarz H.E.}, 2004, A\&A, 417, 637

\bibitem[Feast et al. (2003)]{Feast2003}
      {Feast M.W., Whitelock P.A., Marang, F.}, 2003,
      MNRAS, 346, 878

\bibitem[Fox\ \& Wood (1982)]{FW82}
       {Fox M.W.,  Wood P.R.}.  1982, ApJ, 259, 198

\bibitem[Jones \&\ Gehrz (1990)]{Jones1990}
        {Jones T.J., Gehrz R.D.}, 1990, AJ, 100, 274

\bibitem[Herwig (2005)]{Herwig2005}
     {Herwig F.,} 2005,
     ARA\&A, 43, 435

\bibitem[Herwig (2004)]{Herwig2004}
     {Herwig F.,} 2004,
     ApJS, 155, 651

\bibitem[Icket et al. (1992)]{Icke1992}
      {Icke V., Frank A., Heske A.}, 1992, A\&A, 258, 341

\bibitem[Ita et al. (2004)]{Ita2004}
      {Ita Y., Tanab\'e T., Matsunaga N., et al.}, 2004,
       MNRAS, 347, 720

\bibitem[Kiss \&\ Bedding (2003)]{Kiss2003}
      {Kiss L.L., Bedding T.R.}, 2003, MNRAS, 343,  L79

\bibitem[Magrini et al. (2003)]{Magrini2003}
       {Magrini L., Corradi R.L.M., Greimel R., Leisy P., Lennon D.J.,
       Mampaso A., Perinotto M., Pollacco D.L., Walsh J.R., 
       Walton N.A., Zijlstra, A.A.}, 2003,
       A\&A, 407, 51

\bibitem[Marshall et al. (2004)]{Marshall2004}
       {Marshall J.R., van Loon J.Th., Matsuura M., Wood P.R.,
         Zijlstra A.A., Whitelock P.A.}, 2004,
         MNRAS, 355, 1348

\bibitem[Matsuura et al. (2005)]{Matsuura2005}
        {Matsuura M., Zijlstra A.A., van Loon J.Th., Yamamura I.,
          Markwick A.J., Whitelock P.A., Woods P.M., Marshall J.R.,
           Feast M.W., Waters L.B.F.M.}, 2005,
         A\&A, 434, 691

\bibitem[Mauron \&\ Huggins (2006)]{Mauron2006}
     {Mauron M., Huggins P.}, 2006, A\&A, in press

\bibitem[Nakashima (2005)]{Nakashima2005}
     {Nakashima J.}, 2005, ApJ, 620, 943

\bibitem[Olivier \&\ Wood (2005)]{Olivier2005}
     {Olivier, E.A., Wood P.R.}, 2005, MNRAS, 362, 1396

\bibitem[Olofsson et al. (2000)]{Olofsson2000}
     {Olofsson H., Bergman P., Lucas R., Eriksson K.,
      Gustafsson B., Bieging J.H.,} 2000, A\&A, 353, 583

 \bibitem[Osterbart et al. (2000)]{Osterbart2000}
     {Osterbart R., Balega Y.Y., Bl\"ocker T., Men'shchikov A.B.,
        Weigelt G.}, 2002, A\&A, 357, 169

\bibitem[Sch\"oier at al. (2005)]{Schoier2005}
      {Sch\"oier F.L., Lindqvist M., Olofsson, H.,} 2005,
      A\&A, 436, 633

\bibitem[Simis et al. (2001)]{Simis2001}
      {Simis Y.J.W., Icke V., Dominik C.}, 2001, A\&A, 371, 205

\bibitem[Sloan et al. (2006)]{Sloan2006}
      {Sloan GC., Kraemer K.E., Matsuura M., Wood P.R.,
        Price S.D., Egan M.P.}, 2006, ApJ, in press

\bibitem[Templeton et al. (2005)]{Templeton2005}
      {Templeton M.R., Mattei J.A., Willson L.A.}, 2004,
      AJ, 130, 776

\bibitem[Van Hron et al. (2003)]{VanHorn2003}
      {Van Horn H.M., Thomas J.H., Frank A.; Blackman E.G.},
      2003, ApJ, 585, 983

\bibitem[Van Loon et al. (2005)]{VanLoon2005}
     {van Loon J.Th., Cioni M.-R.L., Zijlstra A.A., Loup C.,} 2005,
     A\&A, 438, 273

\bibitem[Van Loon et al. (1999)]{VanLoon1999}
       {van Loon J.Th., Groenewegen M.A.T., de Koter A., Trams N.R.,
        Waters L.B.F.M., Zijlstra A.A., Whitelock P.A., Loup C.}, 1999,
     A\&A, 351, 559

\bibitem[Vassiliadis \&\ Wood (1993)]{Vassiliadis1993}
     {Vassiliadis E., Wood P.R.,} 1993,
     ApJ, 413, 614

\bibitem[Villaver et al. (2002)]{Villaver2002}
     {Villaver E., Garc\i a-Segura G., Manchado A.}, 2002,
     ApJ, 571, 880

\bibitem[Wachter et al. (2002)]{Wachter2002}
     {Wachter A., Schr\"oder K.-P., Winters J.M., Arndt T.U.. 
      Sedlmayr E.,} 2002, 
     A\&A, 384, 452

\bibitem[Wareing et al. (2006)]{Wareing2006}
      {Wareing C.J., O'Brien T.J., Zijlstra A.A.,
       Kwitter K.B., Irwin J., Wright N., Greimel R., Drew J.E.},
      2006, MNRAS, 366, 387

\bibitem[Whitelock (1994)]{Whitelock1994}
     {Whitelock P.A.,} 1994,
     MNRAS, 270, L15

\bibitem[Whitelock et al. (2006)]{Whitelock2006}
      {Whitelock P.A., Feast M.W., Marang, F., Groenewegen M.A.T.}, 2006,
      MNRAS, in press

\bibitem[Willson, L.A. (2000)]{Willson2000}
     {Willson L.A.,} 2000, ARA\&A, 38, 573

\bibitem[Winters et al. (2000)]{Winters2000}
     {Winters J.M., Le Bertre T., Jeong K.S., Helling Ch., Sedlmayr, E.,}
     2000, A\&A, 361, 641

\bibitem[Woitke (2005)]{Woitke2005}
     {Woitke P.,} 2005,
     A\&A, 433, 1101

\bibitem[Woitke (2006)]{Woitke2006}
     {Woitke P.,} 2006,
     A\&A, in press

\bibitem[Wood et al. (2004)]{Wood2004} 
       {Wood P.R., Olivier E.A., Kawaler S.D.}, 2004,
       ApJ, 604, 800

\bibitem[Wood (1990)]{Wood90} {Wood} P.R., 1990, in From Miras to
Planetary Nebulae: Which Path for Stellar Evolution?, M.O Mennessier
\&\ A. Omont (eds.), (Edition Frontieres, Gif-sur-Yvette), p 67

\bibitem[Ya'ari \&\ Tuchman (1996)]{Yaari1996}
        {Ya'Ari A., Tuchman Y.}, 1996, ApJ, 456, 350

\bibitem[Zijlstra et al. (2001)]{Zijlstra2001}
       {Zijlstra A.A., Chapman J.M., te Lintel Hekkert P., Likkel L.,
         Comeron F., Norris R.P., Molster F.J., Cohen R.J.}, 2001
         MNRAS, 322, 280

\bibitem[Zijlstra et al. (2002)]{Zijlstra2002}
     {Zijlstra A.A., Bedding T.R., Mattei J.A.,} 2002,
     MNRAS, 334, 498

\bibitem[Zijlstra \&\ Bedding (2002)]{Zijlstra2002a}
     {Zijlstra A.A., Bedding T.R.}, 2002, JAVSO, 31, 2

\bibitem[Zijlstra \&\ Weinberger (2002)]{Zijlstra2002b}
     {Zijlstra A.A., Weinberger R.}, 2002, ApJ, 572, 1006

\bibitem[Zijlstra (2004)]{Zijlstra2004}
       {Zijlstra A.A.}, 2004
         MNRAS, 348, L23

\bibitem[Zijlstra et al. (2004)]{Zijlstra2004a}
       {Zijlstra A.A., Bedding T.R., Markwick A.J., Loidl-Gautschy R., et al.},
       2004, MNRAS, 352, 325

\bibitem[Zijlstra et al. (2006)]{Zijlstra2006}
       {Zijlstra A.A., Matsuura M., Wood P.R., Sloan G.C.}, 2006
         MNRAS, submitted

\end{thebibliography}
\end{document}